\documentstyle[twoside,fleqn,npb,epsfig]{article}
\def\Journal#1#2#3#4{{#1} {\bf #2}, #3 (#4)}

\def\NPB{{\em Nucl. Phys.} {\bf B}}
\def\PLB{{\em Phys. Lett.} {\bf B}}
\def\PRL{\em Phys. Rev. Lett.}
\def\PRD{{\em Phys. Rev.}{\bf D}}
\def\ZPC{{\em Z. Phys.} {\bf C}}

\title{A general reduction method for  one-loop N-point integrals\thanks{Talk presented at {\em Loops and Legs in Quantum Field
Theory}, April 2000, Bastei, Germany}}

\author{\underline{G. Heinrich}\address{Laboratoire de Physique Th\'eorique LPT\\
           Universit\'e de Paris XI, B\^atiment 210,
           91405 Orsay, France},
        T. Binoth\address{Laboratoire d'Annecy-Le-Vieux de Physique 
                          Th\'eorique LAPTH\\
                          Chemin de Bellevue, B.P. 110, 74941  
                          Annecy-le-Vieux, France} }

\begin{document}

\begin{abstract}
In order to calculate cross sections with a large number of
particles/jets in the final state at next-to-leading order, 
one has to reduce the occurring scalar and tensor one-loop
integrals to a small set of known integrals. 
In massless theories, this reduction procedure is complicated by the
presence of infrared divergences. Working in  $n=4-2\epsilon$
dimensions, it will be outlined how to achieve such a reduction  
for diagrams with an {\em arbitrary} number of external legs. 
As a result, any integral with more than four propagators and 
generic 4-dimensional external momenta can be reduced to box integrals.
\end{abstract}

\maketitle

\section{Motivation}

The study of processes with multi--particle/jet final states is of major
importance not only to test perturbative QCD, but also to obtain a
precise estimate of the background for  
search experiments. At the
Tevatron, multi--jet cross sections with up to six jets have been
measured and compared to tree level QCD predictions~\cite{abe}. 
However, leading order predictions are very unstable with respect
to variations of the renormalization and factorization scales, such that 
the NLO corrections are necessary to obtain reliable predictions. 

The calculation of e.g. a $2\to N-2$ parton process at NLO requires the
knowledge of $N$-point scalar and tensor one-loop integrals  with
massless propagators and external legs, 
for $e^+e^- \to N\, jets$ one needs $(N+1)$-point integrals with 
one off-shell external leg. 
These integrals have to be further processed by
reducing the tensor integrals to scalar integrals
and, for $N\ge 5$,  reducing the scalar $N$-point integrals to 
known integrals with less propagators.

Of course, reduction methods have been worked out and successfully 
applied before~\cite{Melrose}--\cite{Denner}, 
but mostly for integrals with massive propagators, 
as needed e.g. for electroweak radiative
corrections. In contrast, when dealing with massless partons, 
the presence of infrared divergences requires a different approach.  
For this case, reduction methods have been worked 
out~\cite{Davydychev}--\cite{Weinzierl} which are valid for up to 6 
external legs. 
How to extend these results to more than 6 external legs was not clear until 
recently~\cite{BGH} due to problems stemming from the inversion
of kinematical matrices which were singular if the external momenta were
kept in four dimensions. On the other hand, the use of helicity
techniques requires 4-dimensional external momenta. 
The main ideas how to overcome this problem will be outlined in the
following, leading to reduction formulas for scalar and tensor
integrals valid for an arbitrary number of external legs.

\section{Reduction of  scalar integrals}

\begin{figure}[h]
\begin{center}
    \epsffile{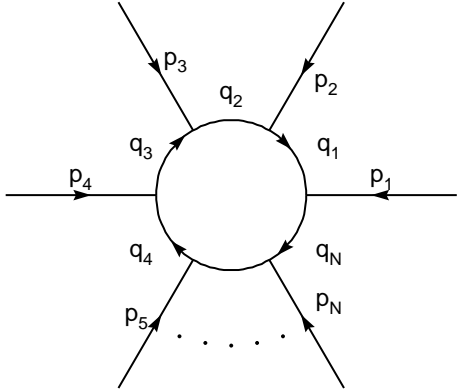}
\caption{\em The one--loop N--point graph.\label{npo-graph}}
\end{center}
\end{figure}

Consider the scalar integral 
\begin{eqnarray*}
I_N^n&=& \int \frac{d^n k}{i\pi^{n/2}} \; \frac{1}{\prod^N_{l=1} q_l^2}\\
q_l &=& k- r_l\quad ,\quad r_l=\sum\limits_{j=1}^l p_j
\end{eqnarray*}
in
$n=4-2\epsilon$ dimensions with $N$ massless propagators, 
depicted in Fig.~\ref{npo-graph}. 

The corresponding expression in  Feynman parameter space reads
\begin{eqnarray*}
I_N^n &=& 
(-1)^N \Gamma(N-n/2) \\ && \times \int_{0}^{\infty} d^Nz 
\frac{\delta(1-\sum_{l=1}^N z_l)}{(z\cdot S \cdot z)^{N-n/2}} \;.
\end{eqnarray*}
The kinematic information is contained in the matrix $S$
which is related to the Gram matrix $G$ by
\begin{eqnarray} \label{EQStoG}
S_{kl}&=&-\frac{1}{2}(r_l-r_k)^2 = \frac{1}{2}( G_{kl} - r^2_l -r^2_k )\\
G_{kl}&=&2\,r_k\cdot r_l \; ,
\quad k,l=1,\ldots,N \;.\nonumber
\end{eqnarray}
The external momenta are kept in four dimensions; momentum conservation 
implies $r_N=0$. 
The integral $I_N^n$ can be split into a singular part with less
propagators and a finite, but formally higher dimensional integral as
\begin{eqnarray}
I_N^n & = & I_{div} + I_{fin}\nonumber\\
I_{div}&= & \int \frac{d^n k}{i\pi^{n/2}} \; 
\frac{\sum^N_{l=1} b_l \, q_l^2}{\prod^N_{l=1} q_l^2}\label{idiv}\\
I_{fin}&=& \int \frac{d^n k}{i\pi^{n/2}} \; 
\frac{\left[ 1 - \sum^N_{l=1}  b_l \, q_l^2 \right] }{\prod^N_{l=1}
q_l^2}\nonumber\\
&=&-\left( \sum_{l=1}^N b_l \right) (N-n-1)\, I_N^{n+2}\;,
\label{EQifin2}
\end{eqnarray}
where the coefficients $b_l$ have to fulfill the following equations:
\begin{eqnarray}\label{EQbl}
\sum\limits_{l=1}^{N}S_{kl} b_l&=&-\frac{1}{2}\quad\Leftrightarrow
\nonumber\\
\sum\limits_{l=1}^{N-1} G_{kl} b_l &=& r^2_k \sum\limits_{l=1}^{N}  b_l
 \quad, \quad 
\sum\limits_{l=1}^{N-1} r^2_{l} b_l = 1 
\end{eqnarray}
The integral $I_{div}$ in (\ref{idiv}) is a sum over reduced integrals
where one propagator could be cancelled. 
Our aim is to show that the higher dimensional integrals $I_N^{n+2}$ in
(\ref{EQifin2}), which are very hard to calculate for $N>4$ and hinder a
reduction, drop out for $N>4$. To
this end we have to solve Eqs.~(\ref{EQbl}) for the parameters $b_l$. 
Note that for $N=5$, expression (\ref{EQifin2}) is of order $\epsilon$, 
which means that  $I_{fin}$
can be dropped right away in that case because 
 $(n+2)$-dimensional integrals are finite for $N\ge 4$.

 Since only four 4-dimensional
external vectors $r_i^{\mu}$ can be linearly independent in Minkowski
space, rank($S$)=min($N$,6) and rank($G$)=min($N$-1,4). 
Hence the case $N=6$ is special since  $S$ is still invertible for $N=6$, 
whereas $G$ is not. Therefore Eqs.(\ref{EQbl}) can still be solved by 
inverting $S$ for $N=6$, whereas for $N>6$ both, $G$ and $S$ have a
vanishing determinant. 
In this case, 
the solution to (\ref{EQbl}) can be constructed in terms of the 
so-called {\em pseudoinverse} $H$ to $G$. It is defined by the 
properties $HGH=H$, $GHG=G$. Given a pseudoinverse $H$ a solution
to $Gx=y$ exists if and only if $y=GHy$.
Introducing a basis $E^{\mu}_{l=1,\dots ,4}$ of Minkowski space 
and defining the matrices $R$ and $\tilde G$ by
\begin{equation}
r_j^{\mu} = \sum\limits_{l=1}^{4} R_{lj} E^{\mu}_l \, , 
\quad\tilde G_{jk} = 2 \,E_j\cdot E_k\; ,
\end{equation} 
the uniquely defined\footnote{In general $H$ is not unique. 
In our case, the uniqueness follows from the symmetry of $G$.} 
pseudoinverse is given by
\begin{equation}
H = R^T \,(R\,R^T)^{-1} \,\tilde G^{-1}\, (R\,R^T)^{-1}\, R\;.
\label{EQdefpseudoinvers}
\end{equation} 
Now Eq.~(\ref{EQbl}) can be solved for general $N$ \cite{BGH}. 
Doing so one finds 
$$\sum^N_{l=1} b_l=0\quad\mbox{ for } N\ge 6$$ 
such that the contribution (\ref{EQifin2}) completely vanishes for all $N\ge 6$.
This means that all scalar $N-$point integrals, for arbitrary $N$, can -- 
by recursion -- be expressed in terms of box integrals. Explicitly, one
finds the following reduction formulas:

For $N\le 6$, i.e. $\det(S)\not = 0$:
\begin{eqnarray}\label{EQscalarreductionNT}
I_N^n &=& -\frac{1}{2}\sum \limits_{k,l=1}^{N} S^{-1}_{kl} I_{N-1,l}^n 
\nonumber\\
&&+ (N-n-1) \,\frac{\det(G)}{2^{N}\det(S)}\, I_N^{n+2} \;.\label{sca1}
\end{eqnarray}
For $N > 6$, i.e. $\det(S) = 0$:
\begin{eqnarray}\label{EQscalarreductionT}
I_N^n &=& \frac{-1}{v\cdot K\cdot v}\sum \limits_{l=1}^{N-1} 
\Bigl( (K\cdot v)_l+\sum_{k=1}^{N-6} \beta_k  U^{(k)}_l\Bigr) \nonumber\\
&&\left(I_{N-1,N}^n -I_{N-1,l}^n\right) \;.\label{sca2}
\end{eqnarray}
The reduced integrals $I_{N-1,l}$ are defined as
\begin{eqnarray}\label{EQreducedgraph}
I^n_{N-1,l} &=&\int \frac{d^n k}{i\pi^{n/2}} \; \frac{ 
 q_{l}^2 }{\prod^N_{k=1} q_k^2}
\nonumber\\
\mbox{and }\quad K&=&1_{N-1}-H\, G\, ,\quad v_l=r_l^2\;.\nonumber
\end{eqnarray}
The $(N-6)\; \beta_k$ in (\ref{EQscalarreductionT}) are free parameters 
to construct linear combinations
of the vectors $\{U^{(k)}\} \,(k=1,\ldots,N-6)$ which, together with
$U^{(N-5)}$, form a basis for the kernel of $G$. 
We chose $$U^{(N-5)}=K\cdot v/(v\cdot K\cdot v)$$ parallel to $v$ and
the others orthogonal, $$v\cdot U^{(k=1,\dots,N-6)}=0 \, .$$ 
Obviously one can choose the $\beta$'s such that 
$(N-6)$ $b$'s from the set 
$\{b_1,\dots , b_{N-1}\}$ are zero.
Doing so one observes that
$I_N^n$ can be expressed by only 6 $(N-1)$-point graphs for arbitrary
$N\ge 6$. This is the generalization to massless kinematics of a similar
result in the infrared finite case~\cite{VanNeervenVermaseren} which has
been derived for integer dimensions using 4-dimensional Schouten
identities. 

\section{Reduction of tensor integrals}

Tensor integrals can also be split into a part containing reduced
integrals ($K_N^{\cdots}$) and a part containing higher 
dimensional objects ($J_N^{\cdots}$). Using momentum conservation
($r_N=0$) one can write:
\begin{eqnarray}
I_N^{{\mu_1}\dots {\mu_L}} &=& \int \frac{d^n k}{i\pi^{n/2}}\; 
\frac{k^{\mu_1}\dots k^{\mu_L}}{\prod^N_{l=1} q_l^2}\nonumber\\
&=&K_N^{{\mu_1}\dots {\mu_L}}+J_N^{{\mu_1}\dots {\mu_L}}\nonumber\\
&&\nonumber\\
K^{\mu_1 \dots \mu_L}_N &=& \sum\limits_{l=0}^{L}\sum 
\limits_{j_1,\dots,j_l=1}^{N-1}  
{\cal F}(\{r_j\})^{\mu_1\dots\mu_L}_{j_1,\dots,j_l}
\nonumber\\
&& \times I^n_{N-l,N-j_1,\dots,N-j_l}  \nonumber\\
J^{\mu_1 \dots \mu_L}_N  &=&  \sum \limits_{l=1}^{[L/2]}  
\sum \limits_{j_1,\dots,j_{L-2l}=1}^{N-1} 
{\cal G}(\{r_j\})^{\mu_1\dots\mu_L}_{j_1,\dots,j_{L-2l}}\nonumber\\
&& \times I_N^{n+2l}(j_1,\dots,j_{L-2l}) 
\label{36b}
\end{eqnarray}
where 
\begin{eqnarray}
I^n_{N-l,N-j_1,\dots,N-j_l} =\int\frac{d^n k}{i\pi^{n/2}}  \frac{ 
\prod\limits^l_{m=1} (q_N^2-q_{j_m}^2) }{\prod^N_{i=1} q_i^2}\nonumber
\end{eqnarray} 
are differences of reduced integrals and 
$$I_N^{n+2l}(j_1,\dots,j_{L-2l})$$ 
are  higher dimensional integrals 
with Feynman parameters $z_j$ in the numerator. 
The Lorentz tensors ${\cal F},{\cal G}$ present in the above
expressions can be written 
in terms of the objects ($v_l=r_l^2$)
\begin{eqnarray}
{\cal H}^{\mu\nu} &=&  r^\mu \cdot H \cdot r^\nu  \nonumber \\
{\cal K}^{\mu}_l &=&  ( r^\mu \cdot H )_l \nonumber \\
{\cal W}^\mu &=&  {\cal K}^{\mu}\cdot v \label{Lorentz}
\end{eqnarray}
Note that the pseudoinverse $H$
to the Gram matrix $G$  also appears here.  
The important feature is that one always has
$${\cal G} \sim g^{\mu\nu}/2-r^\mu \cdot H \cdot r^\nu \, .$$ 
If now the Lorentz vectors $r_j^{\mu}\,(j=1,\ldots ,N-1)$ 
span Minkowski space, which is generically the case for $N\ge 5$, 
one immediately derives, using the explicit expresssion of $H$ given
above,  
that $$r^\mu \cdot H \cdot r^\nu=(g^{\mu\nu}/2)^{4-{\rm dim}}\, .$$ 
This means 
$$g^{\mu\nu}/2-r^\mu \cdot H \cdot r^\nu ={\cal O}(\epsilon)$$ and thus
the higher dimensional objects $I_N^{n+2l}$ in Eq.~(\ref{36b}) can be dropped for
$N\ge 5$. Consequently it follows from Eq.~(\ref{36b})  that for the
reduction of  
an $N$-point integral only a  small number of higher
dimensional integrals is needed, i.e. one has to know $J_N^{\mu_1\dots \mu_L}$
only for $2\le L\le N=2,3,4$. These can be calculated  once and forever, 
the explicit expressions are 
\begin{eqnarray}\label{EQhigherdimterms}
J^{\mu_1\mu_2}_{N=2,3,4} = -\left( g^{\mu_1\mu_2}/2 - 
{\cal H}^{\mu_1\mu_2} \right) I_N^{n+2}\qquad\;\;\nonumber\\
J^{\mu_1\mu_2\mu_3}_{N=3,4}\, =  
-\Bigl[ \left( g/2 - {\cal H} \right)^{\cdot\cdot} 
{\cal W}^{\cdot}  \Bigr]^{\{\mu_1\mu_2\mu_3\}} I_N^{n+2} 
\nonumber\\
- \Bigl[ \left( g/2 - {\cal H} \right)^{\cdot\cdot} 
{\cal K}_l^{\cdot}\Bigr]^{\{\mu_1\mu_2\mu_3\}} I_{N-1,N-l}^{n+2}
\nonumber\\
J^{\mu_1\mu_2\mu_3\mu_4}_4 =\Bigl[\left( g/2-{\cal H}\right)^{\cdot\cdot}\qquad\qquad\qquad\qquad\nonumber\\ 
\times\left( g/2 - {\cal H} \right)^{\cdot\cdot} \Bigr]^{\{\mu_1\mu_2\mu_3\mu_4\}} 
I_4^{n+4}\nonumber\\
- \Bigl[ \left( g/2 - {\cal H} \right)^{\cdot\cdot} 
{\cal  W}^{\cdot}{\cal  W}^{\cdot}\Bigr]^{\{\mu_1\mu_2\mu_3\mu_4\}} 
I_4^{n+2}\nonumber\\
- \frac{1}{2}\Bigl[ \left( g/2 - {\cal H} \right)^{\cdot\cdot} 
{\cal  W}^{\cdot}
{\cal  K}_l^{\cdot}\Bigr]^{\{\mu_1\mu_2\mu_3\mu_4\}} 
I_{3,4-l}^{n+2}\nonumber\\
- \Bigl[ \left( g/2 - {\cal H} \right)^{\cdot\cdot} 
{\cal H}^{\cdot}_{\nu} {\cal  K}^{\cdot}_l\Bigr]^{\{\mu_1\mu_2\mu_3\mu_4\}} 
I_{3,4-l}^{\nu,n+2}\;.\nonumber
\end{eqnarray} 
The indexed bracket is an abbreviation for the sum of 
all distinguishable distributions
of Lorentz indices to the objects inside the bracket, e.g.
\begin{eqnarray}
\Bigl[ X^\cdot Y^\cdot Y^\cdot \Bigr]^{\{\mu_1\mu_2\mu_3\}}=
X^{\mu_1}Y^{\mu_2}Y^{\mu_3} \qquad\qquad\nonumber\\
+ X^{\mu_2}Y^{\mu_3}Y^{\mu_1}
+X^{\mu_3}Y^{\mu_1}Y^{\mu_2} .\nonumber
\end{eqnarray}
For the $n$-dimensional part, $K_N^{\mu_1\dots \mu_L}$,
a recursion relation holds.  
Skipping all the details (see \cite{BGH}) one finds
\begin{eqnarray}\label{rec}
K_N^{\mu_1\ldots\mu_L}&=&\frac{1}{L} \Bigl[{\cal W}^{\cdot} 
K_N^{\{L-1 \;\mathrm{dots}\}}\Bigr]^{\{\mu_1\ldots\mu_L\}}\nonumber\\
&+&\frac{2^{(L-1)}}{L!}
\Bigl[{\cal K}_l^{\cdot}{\cal H}^{\cdot}_{\nu_1}
\ldots{\cal H}^{\cdot}_{\nu_{L-1}}\Bigr]^{\{\mu_1\ldots\mu_L\}}
\nonumber\\ && \quad \times
I_{N-1,N-l}^{\nu_1\ldots\nu_{L-1}}\;.
\end{eqnarray}

Putting everything together, the final recursion formula for
$N$-point tensor integrals reads
\begin{eqnarray}
I_N^{\mu_1\ldots\mu_L}&=& 
 \frac{1}{L}
\Bigl[{\cal W}^{\cdot}(I_N-J_N)^{\{L-1 \;\mathrm{dots}\}}
\Bigr]^{\{\mu_1\ldots\mu_L\}}\nonumber\\
&&+\frac{2^{(L-1)}}{L!}
\Bigl[{\cal K}_l^{\cdot}{\cal H}^{\cdot}_{\nu_1}\ldots{\cal H}^{\cdot}_{\nu_{L-1}}
\Bigr]^{\{\mu_1\ldots\mu_L\}}
\nonumber\\&& \quad\times
I_{N-1,N-l}^{\nu_1\ldots\nu_{L-1}} + J_N^{\mu_1\ldots\mu_L}\,.
\label{recwithJ}
\end{eqnarray}
This form is not immediately suited for iteration, as some
reduced integrals do not contain the trivial propagator $k^2$ 
anymore. Therefore one has to shift the loop momenta
of the respective integrals by $k\to k+r_l$ before iteration,
leading  to 
 \begin{eqnarray}\label{EQshift}
&&I^{\mu_1\dots\mu_P}_{N-1,N-l}(R)
=\nonumber\\
&&  I^{\mu_1\dots\mu_P}_{N-1}(\hat R_{[N]})
-I^{\mu_1\dots\mu_P}_{N-1}(\hat R_{[l]})\nonumber\\
&&= \sum\limits_{k=0}^{P} \Bigl[ r_{l}^{\cdot (k)} 
I_{N-1}^{\{P-k
\;\mathrm{dots}\}}(R_{[l]})\Bigr]^{\{\mu_1\dots\mu_{P}\}}\nonumber\\ 
&&-I_{N-1}^{\mu_1\dots\mu_{P}}(\hat R_{[l]})\;.
 \end{eqnarray}
The argument vectors are
\begin{eqnarray}\label{EQargumentvectors}
R &=& (r_1,\dots,r_{N}) \nonumber\\
\hat R_{[k]} &=& (r_{1},\dots,\hat r_{k} ,\dots , r_N)\nonumber\\
R_{[l]}   &=& (r_{l+1}-r_{l},r_{l+2}-r_{l},\dots,r_{N-2+l}-r_{l},0)\nonumber
\end{eqnarray}
where $\hat r$ means that the respective vector is missing.
The vector indices are understood to be taken cyclically symmetric
with periodicity $N$, i.e.
$r_{N+l}=r_{l}$ for $l\in \{1,\dots,N\}$.
As we assume $r_N=0$, the integrals on the right-hand side  of (\ref{EQshift})
have again at least one trivial propagator and thus are 
suited for a further reduction step. As a simple example, consider the 
rank two 3-point integral $I^{\mu_1\mu_2}_3(R)$. After the first
reduction step and subsequent shift, one obtains
\begin{eqnarray}
I^{\mu_1\mu_2}_3(R) 
=\frac{1}{2}\Bigl({\cal W}^{\mu_1} \, I_3^{\mu_2}(R)+{\cal W}^{\mu_2} \, I_3^{\mu_1}(R)\Bigr)\nonumber\\
+\sum\limits_{l=1}^{2}\left({\cal K}^{\mu_1}_l {\cal H}^{\mu_2}_{\nu}+{\cal K}^{\mu_2}_l {\cal H}^{\mu_1}_{\nu}\right)\nonumber\\
\times\left(r_l^{\nu}I_2^n(R_{[l]})+I_2^{\nu}(R_{[l]})-I_2^{\nu}(\hat R_{[l]})\right)\nonumber\\
-(g^{\mu_1\mu_2}/2-{\cal H}^{\mu_1\mu_2})I_3^{n+2}(R)\nonumber
\end{eqnarray}
The recursive structure together with the shift 
can be  implemented in an algebraic manipulation program
in a straightforward manner. The recursion stops if all
tensor integrals are transformed into expressions
containing scalar integrals and Lorentz tensors composed  of the 
objects given in Eq.~(\ref{Lorentz}). Starting with an $N$-point tensor
integral one finds a linear combination of scalar integrals
with $K\le N$ legs. To these scalar integrals
the reduction formulas (\ref{sca1}),(\ref{sca2}) can be applied.
In this way our reduction formalism solves
the problem of calculating massless $N$-point Feynman diagrams
for arbitrary $N$. The generalization to the case where
some of the propagators are massive will 
work similarly.       

\section{Summary/Outlook}

In conclusion, massless scalar and tensor one-loop 
integrals for graphs with an {\em arbitrary} number of external legs ($N>4$)
can recursively be reduced to linear combinations of (known) box
integrals. 
To deal with the problem of singular Gram matrices we
introduced the concept of the pseudoinverse,
which renders the whole formalism 
mathematically well defined for arbitrary $N$.  The recursive structure
can easily be implemented in algebraic manipulation
programs.
This allows in principle for the NLO calculation of 
multi--jet cross sections with a large 
number of final states. Applications relevant for the near future are 
e.g. the NLO calculations for $pp\to b\bar b\, b\bar b$ 
or  $pp\to 4\;jets/photons$. The most complicated 
analytical structure in such a calculation is the scalar 6-point
function since,  as shown above, 
no dangerous higher-dimensional objects have to be computed.
With the given methods, an explicit 
analytical expression for the  scalar 6-point function  has been 
derived and presented in \cite{BGH}.

\section*{Acknowledgments}

We would like to thank the organizers for their work.
It was a pleasure for  us to  take part in the 
{\em Loops and Legs in Quantum Field Theory 2000} conference. 
This work 
was supported by the EU Fourth Training Programme  
''Training and Mobility of Researchers'', Network ''Quantum Chromodynamics
and the Deep Structure of Elementary Particles'',
contract FMRX--CT98--0194 (DG 12 - MIHT).

\end{document}